\documentclass[aps,pre,twocolumn,superscriptaddress,floatfix]{revtex4}
\usepackage{graphicx}
\usepackage{bm}
\usepackage{amssymb}

\begin{document}

\title{Control of particle clustering in turbulence by polymer additives}

\author{F. De Lillo}
\author{G. Boffetta}
\affiliation{Department of Physics and INFN, University of Torino, 
via P. Giuria 1, 10125 Torino, Italy}
\affiliation{International Collaboration on Turbulence Research}
\author{S. Musacchio} 
\affiliation{CNRS, Laboratoire J.A. Dieudonn\'e UMR 6621 - Parc Valrose, 06108 Nice, France, EU}
\affiliation{International Collaboration on Turbulence Research}

\begin{abstract}

We study the clustering properties of inertial particles in a turbulent
viscoelastic fluid. The investigation is carried out by means of direct
numerical simulations of turbulence in the \mbox{Oldroyd-B} model. The effects of
polymers on the small scale properties of homogeneous turbulence are considered in relation with their consequences on clustering of particles, both lighter and heavier than the carrying fluid. We show that, depending on particle and flow parameters, polymers can either increase or decrease clustering.
\end{abstract}

\maketitle


Clustering of inertial particles in turbulent flows is relevant for meteorology
and engineering, as well as fundamental research.  It is believed to play a
crucial role in rain-drop formation \cite{falkovich_nature02}, as well as in the aggregation of
proto-planetesimals in Keplerian accretion disks \cite{bracco_pof99}. The physical mechanism which
originates such clustering is indeed rather simple: particles heavier than the
fluid in which they are transported experience inertial forces which expel
them from vortices; particles lighter than the fluid are attracted into
vortical structures, for similar reasons \cite{Squires1991, cencini_jot06, Bec2005}.  In realistic flows, however,
particles are advected by the small scale vortical structures of turbulent
flows: these have highly non-trivial statistical features, resulting in a
complex clustering process which is still far from being completely understood.
From the point of view of applications, the properties of concentration and
distribution of inertial particles play a crucial role in engineering and
for the design of industrial processes involving combustion and mixing \cite{Warnatz2006,Rouson2001, Sbrizzai2006}. Suspensions of particles in viscoelastic fluids are used in many products of commercial and industrial relevance \cite{barnes2003}.

In this paper we investigate, by means of direct numerical simulations 
of a turbulent flow, how the clustering properties of a dilute 
suspension of inertial particles can be affected by the addition of 
small amounts of polymer additives. 
The effects induced by polymers on turbulent flows 
are themselves of enormous relevance. It is enough to mention the celebrated 
drag reduction effect which occurs in pipe flows \cite{Lumley1969},
or the recently discovered elastic turbulence regime \cite{gs_nature00}.
Polymers have striking effects also on Lagrangian properties of the flow. 
In particular it has been shown that polymer addition in turbulent flows 
reduces the chaoticity of Lagrangian trajectories \cite{bcm_prl03} 
and affects acceleration of fluid tracers \cite{cmxb_njp08}.
Conversely in the elastic turbulence regime polymers are 
able to generate Lagrangian chaos in flows at vanishing 
Reynolds number, which would be non chaotic in the Newtonian case
\cite{gs_nature01,bcm_prl03}.

\begin{figure}
\includegraphics[width=4.2cm]{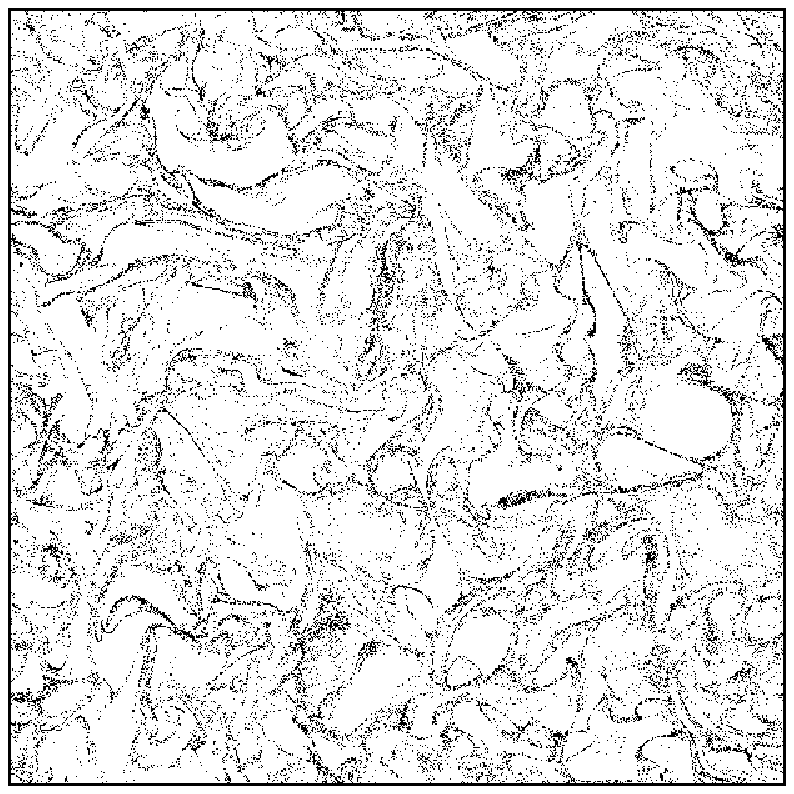}
\includegraphics[width=4.2cm]{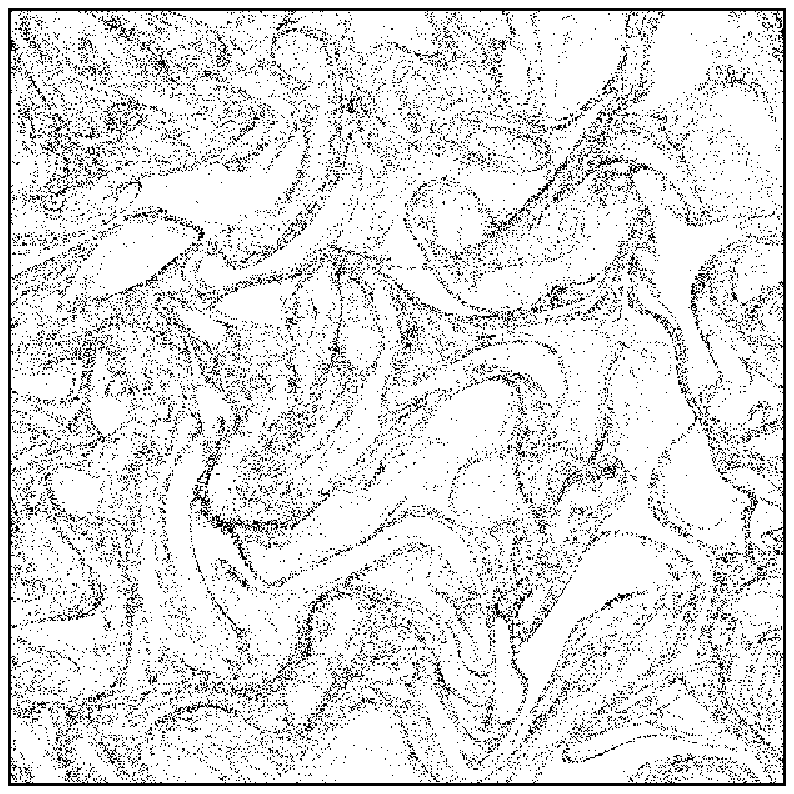}
\includegraphics[width=4.2cm]{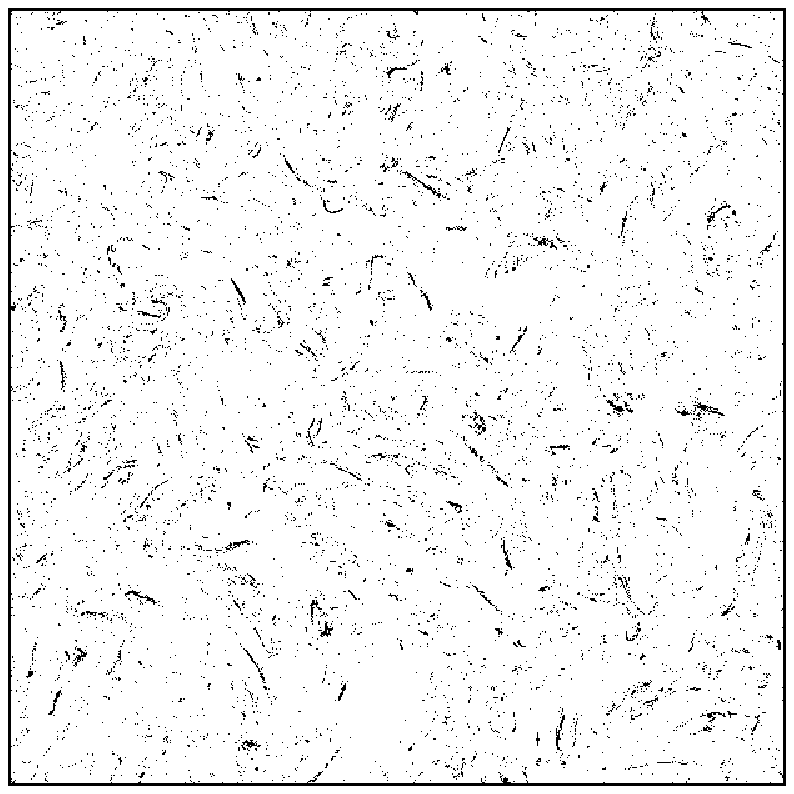}
\includegraphics[width=4.2cm]{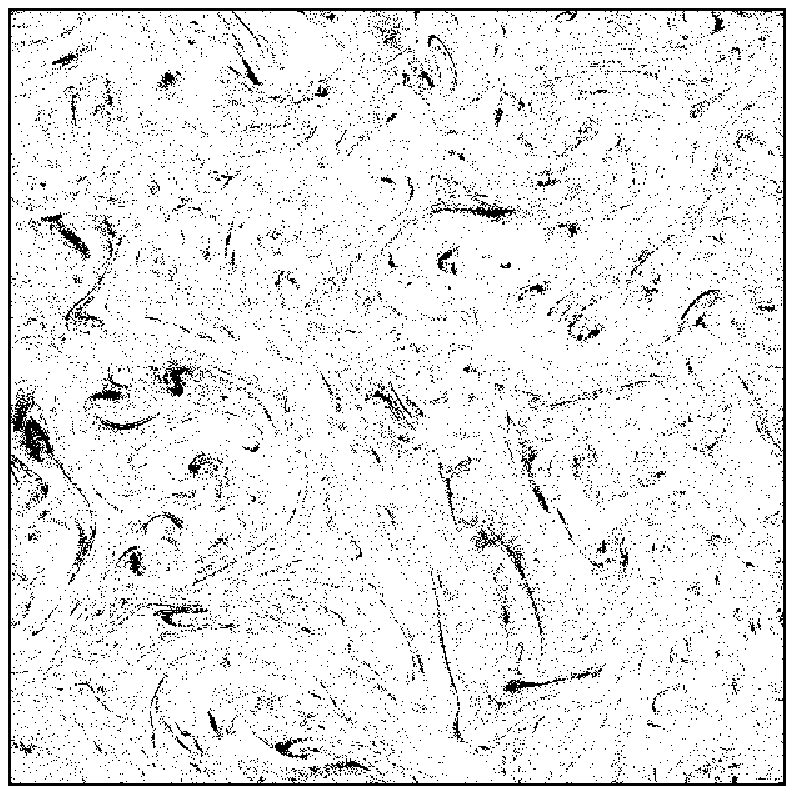}
\caption{Section on plane $z=0$ of the distribution of heavy particles
with $\tau_S=0.035$ (upper panels) and light particles with $\tau_S=0.03$ (lower panels) in statistically stationary conditions in a Newtonian 
flow (left) and a viscoelastic flow at $\rm Wi=1$ (right).
Both flows are forced with the same 
forcing ${\bf f}({\bf x},t)$ $\delta$-correlated in time and localized
on large scales. Numerical simulations are done by a pseudo-spectral,
fully dealiased code at resolution $256^3$. For the viscoelastic simulations,
a small diffusive term is added to (\ref{eq:4}) to prevent numerical
instabilities \cite{sb_jnnfm95}.}
\label{fig1}
\end{figure}

Here we show that the addition of polymers in a turbulent flow has 
important effects on the statistical properties of inertial particles 
which can result in both an increase or a decrease of the clustering.
An example of the effect of polymers on clustering 
is shown in Fig.~\ref{fig1} which represents the distribution 
of an ensemble of inertial particles in a turbulent flow
before and after the introduction of polymers. It is evident, already
at the qualitative level of Fig.~\ref{fig1}, that polymers are able
to change the statistical distribution of particles.
We show that these effects can be understood and quantified 
in terms of the Lyapunov exponents of inertial particles, 
which are very sensitive to the presence of polymers.
Previous systematic investigations of inertial particle dynamics 
in Newtonian turbulent flows \cite{calzavarini_jfm08} 
and stochastic flows \cite{bec_pof03}
have shown that clustering (quantified by means of the Lyapunov Dimension of particle attractor) is maximum when the particle relaxation time
is of the order of the shortest characteristic time of the flow.

We consider the case of a dilute suspension of small inertial particles, 
in which the effects of the disturbance flow induced by the particles 
can be neglected. The dynamics of the suspension is hence 
modeled by an ensemble of non-interacting point particles, 
which experience viscous drag and added mass forces. 
The equation of motion of each particle reads \cite{maxey_pof83}:  
\begin{eqnarray}
{d {\bm x} \over dt}&=& \bm v 
\label{eq:1} \\
{d {\bm v} \over dt}&=&-{1 \over \tau_S}\left[\bm v - \bm u(\bm x(t),t)\right]+
\beta {d {\bm u} \over dt}
\label{eq:2}
\end{eqnarray}
where $\tau_S=a^2/(3\beta\nu)$ is the Stokes relaxation time,
$a$ is the particle radius, $\beta=3\rho_f/(\rho_f+2\rho_p)$ 
($\rho_p$ and $\rho_f$ representing particle and fluid densities
respectively) and $\nu$ is the kinematic viscosity of the fluid 
(replaced by the total viscosity $\nu_T$ in a viscoelastic fluid, see below).
Light (heavy) particles correspond to $\beta>1$ ($\beta<1$). 
In this work we consider the two extreme cases of
very light particles (e.g. air bubble in water) for which $\beta=3$
and very heavy particles with $\beta=0$. 
We define the Stokes number as $\rm St=\tau_S \lambda^ 0_1$, where
$\lambda^0_1$ is the maximum Lyapunov exponent of neutral Lagrangian tracers
(i.e. $\rm St=0$ particles) in the flow. With this definition, 
maximum clustering is obtained for $\rm St \simeq 0.1$ 
\cite{bec_pof03, calzavarini_jfm08}.

The viscoelastic flow ${\bf u}({\bf x},t)$ in which the particles are 
suspended can be described by standard viscoelastic models,
such as the Oldroyd-B model or the nonlinear FENE-P model, 
which accounts for the finite extensibility of polymers. 
In spite of their simplicity, these models are able to reproduce
many relevant properties of dilute polymer solutions, including
turbulent drag reduction~\cite{sbh_pof97,bcm_pre05} and elastic turbulence
phenomenology~\cite{bbbcm_pre08}. 
Here we choose the Oldroyd-B model~\cite{bird87}, in which the 
coupled dynamics of the velocity field ${\bf u}({\bf x},t)$ 
and the polymer conformation tensor $\sigma({\bf x},t)$
(which is proportional to local square polymer elongation) 
reads: 
\begin{eqnarray}
{\partial \bm u \over \partial t}+\bm u\cdot\bm\nabla\bm u &=&
-\bm\nabla p+\nu\nabla^2\bm u+{2\nu\gamma \over \tau_p}\bm\nabla\cdot\sigma
+{\bm f}
\label{eq:3}\\
{\partial \sigma \over \partial t}+\bm u\cdot\nabla\sigma &=&
(\nabla\bm u)^T\cdot\sigma+\sigma\cdot(\nabla\bm u)-
{2 \over \tau_p}(\sigma-\mathbb{I})
\label{eq:4}
\end{eqnarray}
The total viscosity of the solution $\nu_T=\nu (1+\gamma)$ 
is written in terms of the kinematic viscosity of the solvent $\nu$
and the zero-shear contribution of the polymer $\gamma$ which 
is proportional to the polymer concentration.
The polymer time $\tau_p$ represents the longest relaxation time to the equilibrium configuration ($\sigma=\mathbb{I}$ in dimensionless units).
Viscoelasticity of the turbulent flow is parametrized by the 
Weissenberg number $\rm Wi$, the ratio between $\tau_p$ and a characteristic
time of the flow. Here we use $\rm Wi=\tau_p \lambda_1^N$ 
where $\lambda_1^N$ is the Lagrangian Lyapunov exponent of the Newtonian flow,
before the addition of polymers (i.e. (\ref{eq:3}) with $\gamma=0$). We
stress that $\lambda_1^0$ introduced above refers instead to the specific flow
that carries the suspension and it clearly depends on $\rm Wi$. 
Therefore $\lambda_1^N\equiv\lambda_1^0|_{Wi=0}$.

\begin{table}[b]
\label{tab:1}
\begin{tabular}{c c c c c}
$\rm Wi$	&$\varepsilon_f$& $\varepsilon_\nu$&$u_{\rm rms}$	&$\lambda^0_1$	\\
\hline
0	&	0.28	&	0.28	&	0.76	&	1.36	\\
0.5	&	0.28	&	0.18	&	0.73	&	1.08	\\
1	&	0.28	&	0.092	&	0.68	&	0.75	\\
\end{tabular}
\caption{Parameters for the Newtonian and viscoelastic simulations. 
The Weissenberg number $\rm Wi$, energy input $\varepsilon_f$, viscous
dissipation rate $\varepsilon_\nu$, rms velocity $u_{rms}$ and Lagrangian 
Lyapunov exponent $\lambda_1^0$ of the carrier flow are shown. 
In both viscoelastic runs an additional dissipative term was added on 
polymers (see text), with coefficient $\nu_p=2.3\times 10^3$ }
\label{table1}
\end{table}

In the following we discuss results obtained by integrating numerically
the viscoelastic model (\ref{eq:3}-\ref{eq:4}) at high resolution for different 
values of $\rm Wi$ (see Table~\ref{table1}). The flow is sustained by a 
stochastic Gaussian forcing ${\bf f}({\bf x},t)$ 
$\delta$-correlated in time and localized on large scales. Fluid equations were integrated by means of a standard, fully dealiased, pseudo spectral code, on a cubic, triple-periodic domain with 256 grid points per side.
When the flow reaches a turbulent, statistically stationary state, different 
families (i.e. with different values of parameters $\beta$ and $\tau_S$) 
of inertial particles are injected, with initial homogeneous 
distribution in space, and their motion integrated according to 
(\ref{eq:1}-\ref{eq:2}). For each value of $\rm Wi$, we 
integrated the motion of $1024$ particles for each of $21$ values
of $\tau_S$ and two values of $\beta$, namely very heavy particles with 
$\beta=0$ and "bubbles" with $\beta=3$. 

As an effect of inertia the distribution of particles does not remain 
homogeneous and evolves to a fractal set dynamically evolving with 
the flow, such as the examples shown in Fig.~\ref{fig1}. In the language
of dynamical systems, the equations (\ref{eq:1}-\ref{eq:2}) for particle motion
represent a dissipative system whose chaotic trajectories evolve to 
a fractal attractor (which evolves in time following the flow).
A quantitative measure of clustering at small scales 
is therefore obtained by measuring the fractal dimension of the attractor 
(for each family of particles) using the Lyapunov dimension 
\cite{bec_pof03,bec_pof06} defined in terms of Lyapunov exponents as 
$D_L=K+\sum_{i=1}^{K} \lambda_i/|\lambda_{K+1}|$ where $K$ is the 
largest integer for which $\sum_{i=1}^{K} \lambda_i \ge 0$ \cite{ccv2010}.
Since the space distribution of the particles is the projection 
of the attractor on the sub-space of particle positions, 
the fractal dimension of clusters 
is given by $\min(D_L, 3)$~\cite{sy97,hk97}, 
provided that the projection is generic 
(for a discussion on this issue see e.g. \cite{bch07}). 
This implies that $D_L<3$ gives fractal 
distributions of dimension $D_L$, while $D_L>3$ corresponds to space-filling
configurations, which however can be non-homogeneous. 

\begin{figure}
\includegraphics[width=8.0cm]{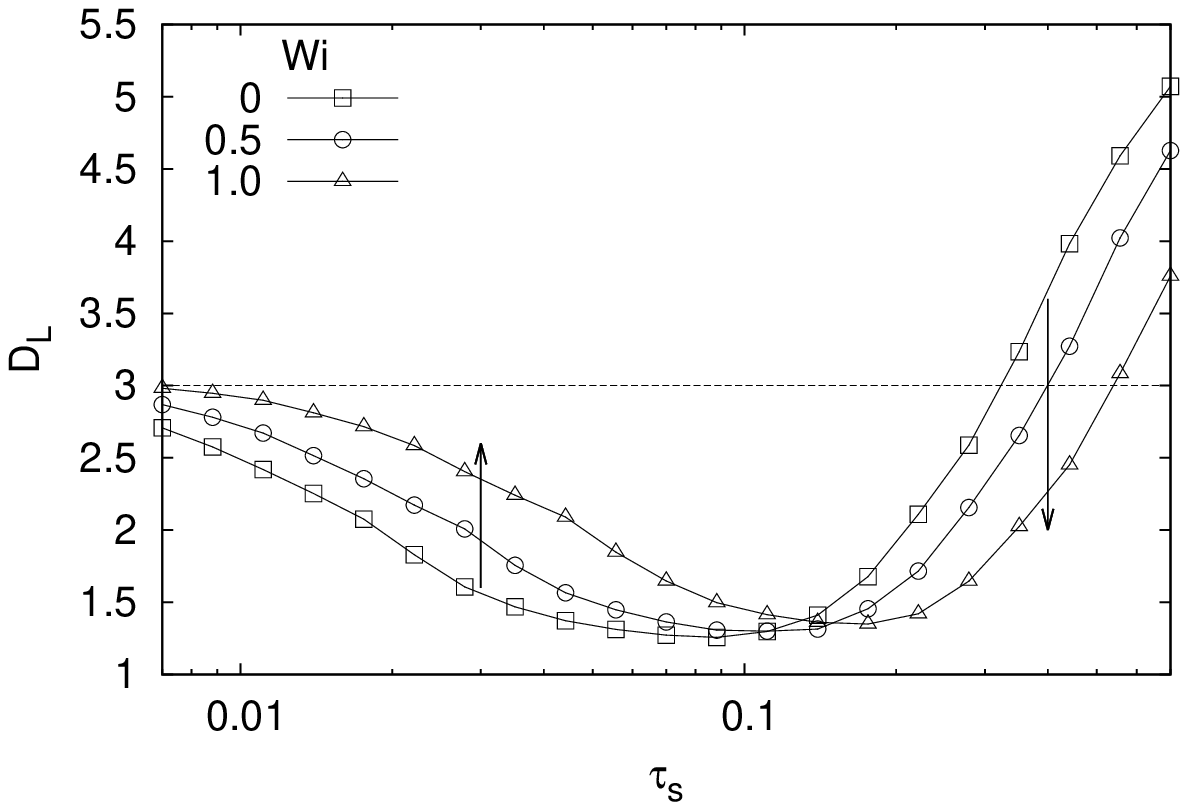}
\includegraphics[width=8.0cm]{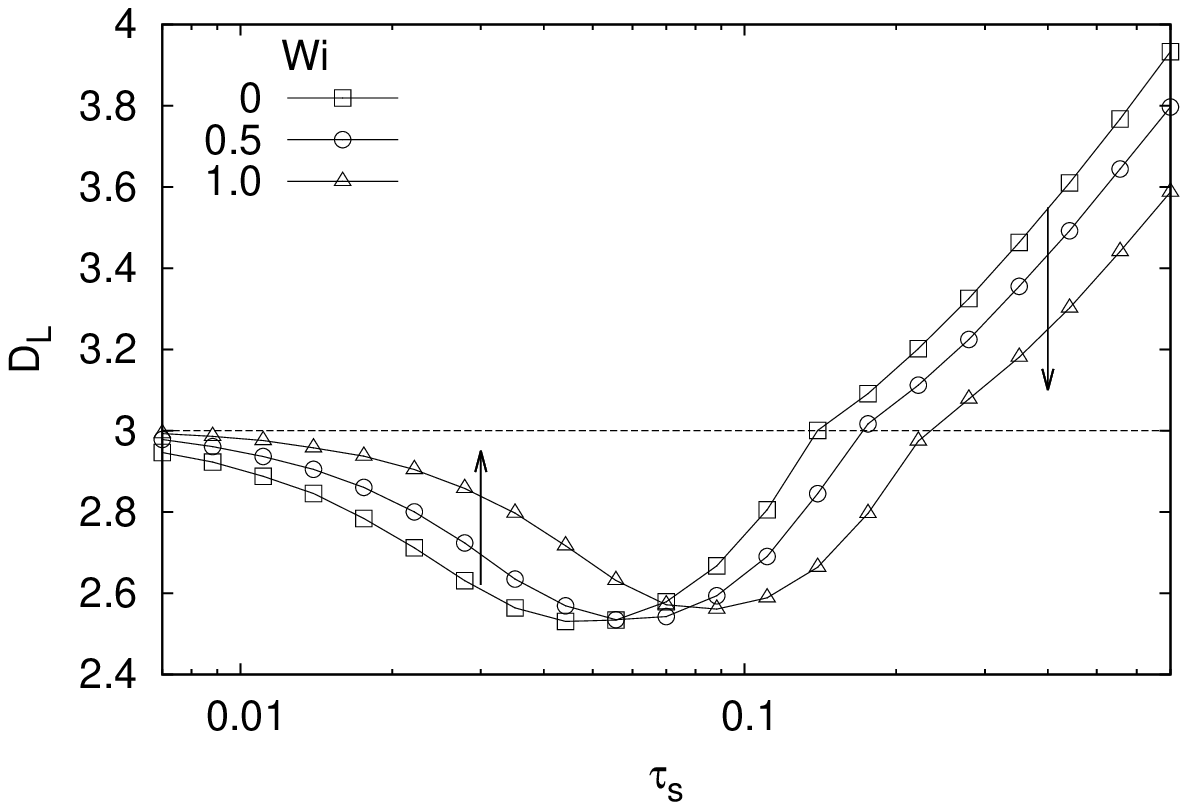}
\caption{
Lyapunov dimension for light (upper panel) and heavy (lower panel) particles
plotted as a function of $\tau_S$. Different lines correspond
to the different Weissenberg numbers: $\rm Wi=0$ (squares), $\rm Wi=0.5$ (circles) and
$\rm Wi=1.0$ (triangles). 
}
\label{fig2}
\end{figure}

In Fig.~\ref{fig2} we plot the fractal dimensions for both heavy
and light particles as a function of $\tau_S$ for the three simulations 
at different $\rm Wi$.
It is evident that the addition of polymer changes substantially 
the clustering properties of the particles, both increasing 
$D_L$ and reducing $D_L$ depending on value of $\tau_S$. 
Figure~\ref{fig1} shows examples of clustering reduction, for heavy and light particles respectively. The upper panels refer to heavy particles ($\beta=0$) with $\tau_S=0.035$, while the bottom ones are extracted from a simulation with $\beta=3$ and $\tau_S=0.03$. Both values of Stokes time are, for the Newtonian flow, on the left of the minimum in $D_L$. As a consequence, polymers produce a reduction of clustering. Such effect is more visible for light particles. A possible reason for this difference will be discussed further on.

\begin{figure}
\includegraphics[width=9.0cm]{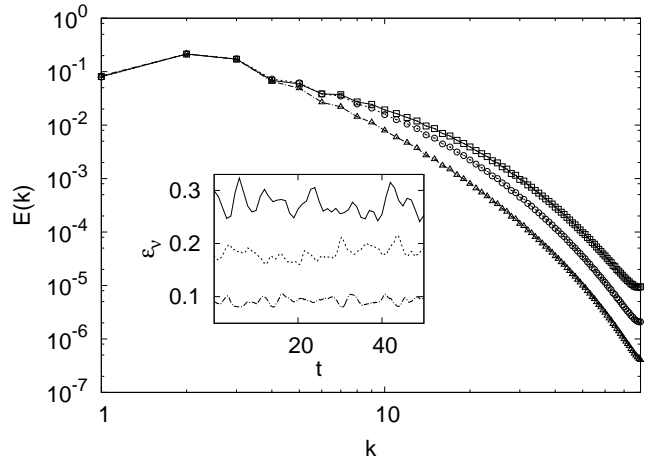}
\caption{
Energy spectra for the Newtonian case ${\rm Wi}=0$ (squares) and for the
viscoelastic ones ${\rm Wi}=0.5$ (circles) and ${\rm Wi}=1$ (triangles). The
depletion due to polymer feed-back is evident on large wavenumbers, while the
larger scales are unaffected. The effect of polymers extends at lower
wave-numbers as $\rm Wi$ increases. Inset: viscous energy dissipation $\varepsilon_\nu$ during
a typical time interval in the stationary simulations, for the Newtonian (solid
line), ${\rm Wi}=0.5$ (dashed line) and ${\rm Wi}=1$ (dash-dot) flows. The
decrease in $\varepsilon_\nu$ with ${\rm Wi}$ is evident, as well as the reduction in
fluctuations.
}
\label{fig3}
\end{figure}

The mechanism at the basis of this effect is not trivial and
is a consequence of the change induced by the polymers on the 
small-scale properties of the turbulent flow. 
In Fig.\ref{fig3} we plot the energy spectra for the different $\rm Wi$ numbers.
The effect of polymers is evident in the high-wavenumber range 
where velocity fluctuations are clearly suppressed, resulting in a 
depletion of the energy spectrum, while large-scale fluctuations are 
unaffected.

Indeed one can expect that only the fastest eddies of the flow, 
i.e. those whose eddy turn-over time $\tau_\ell$ is shorter that 
the polymer relaxation time $\tau_p$, 
can produce a significant elongation of polymers. 
The elastic feedback therefore affects only small scales $\ell$ 
with $\tau_\ell < \tau_p$. 
Conversely, large scales exhibit the same phenomenology of a 
Newtonian flow, characterized by a turbulent cascade with a 
constant energy flux equal to the energy input rate $\varepsilon_f$. 
The turbulent cascade proceeds almost unaffected by the presence 
of polymers down to the Lumley scale $\ell_L$,
whose eddy turn-over time equals the polymer relaxation time. 
A dimensional estimate, based on the Kolmogorov scaling for 
the typical velocity $u_\ell\sim\varepsilon_f^{1/3}\ell^{1/3}$ 
and turn-over time $\tau_\ell=\ell/u_\ell \sim\varepsilon_f^{-1/3}\ell^{2/3}$ 
of an eddy of size $\ell$, gives $\ell_L=\tau_p^{3/2}\varepsilon_f^{1/2}$. 
Polymers would therefore affect only the small scales $\ell < \ell_L$.
Our results are in qualitative agreement with this picture: the
$\rm Wi=0.5$ spectrum differs from the Newtonian 
one only for $k\gtrsim 8$, while
at $\rm Wi=1$ polymers are active over a larger range of scales.
The reduction of kinetic energy at small scales, due to the transfer of 
energy to the polymers, is accompanied by a reduction of the viscous 
dissipation $\varepsilon_\nu=\nu\langle (\nabla u)^2\rangle$ 
at fixed energy input $\varepsilon_f$, 
as can be seen from Table~\ref{table1} and in the inset of 
Fig.\ref{fig3}. 
This phenomenon has been previously observed both in forced and decaying
simulations of statistically homogeneous and isotropic turbulence
(see, e.g., ~\cite{dcbp05,pmp06}). 

\begin{figure}
\includegraphics[clip=true,keepaspectratio,width=8.0cm]{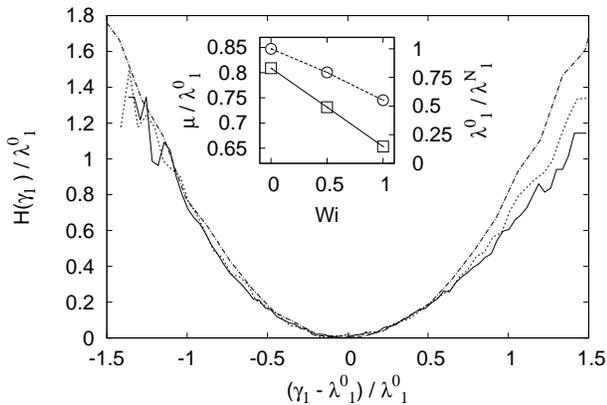}
\caption{Comparison between the Cram\'er functions of the stretching rate
$\gamma_1$ computed at ${\rm Wi}=0$ (solid line),${\rm Wi}=0.5$ (dashed
line),${\rm Wi}=1$ (dash-dot). Inset: first Lagrangian Lyapunov exponent
$\lambda^0_1$ (circles) and width $\mu$ (squares) of the Cram\'er function (see
text) as a function of $\rm Wi$. The Lyapunov exponents are compared with the
Newtonian value $\lambda_1^N$. 
}
\label{fig4}
\end{figure}

The suppression of small-scale motions caused by polymers 
has major consequences also on the Lagrangian statistics. 
It is responsible of the reduction of chaoticity 
of Lagrangian trajectories~\cite{boffetta_prl03}.
Indeed the chaoticity of the flow is directly related to its 
stretching efficiency via the Lyapunov exponents. 
When polymers are stretched, the elastic stress tensor produces a negative
feed-back on small scale stretching, thus reducing the degree of chaoticity of
the flow \cite{boffetta_prl03,balkovsky_pre01}. This effect is clearly
observable in the decrease of the Lagrangian Lyapunov exponent of the flow 
at increasing polymer elasticity (see the inset of Fig.\ref{fig4}). 

It is worth to notice that, because of polymers counteraction, 
the Lyapunov exponent of the resulting viscoelastic flow is smaller than $\tau_p^{-1}$. 
In other words, the $\rm Wi$ number computed a posteriori (i.e. after polymer injection) 
is always smaller than unity. 
This is not in contrast with the hypothesis that polymers have a strong active effect on the flow 
mainly when they are stretched, i.e. above the so-called coil-stretch transition, which is
expected to happen around $\rm Wi\simeq 1$ \cite{balkovsky_prl00}.
Indeed, the Lyapunov exponent simply provides a measure of the 
average stretching in a chaotic flow. One should bear in mind that large fluctuations 
of the stretching rates (and therefore strong viscoelastic effects) 
can occur also when $\rm Wi \lesssim 1$. 

Detailed information on the fluctuations of the stretching rates can be 
obtained from the statistics of the Finite Time Lyapunov Exponents (FTLE)
$\gamma_i$.  
The FTLE are defined via the exponential growth rate during a finite time
$T$ of an infinitesimal $M$-dimensional volume as
$\sum_{i=1}^M\gamma_i=(1/T)\ln[V^M(T)/V^M(0)]$ \cite{ccv2010}.
From the definition of the Lyapunov exponents it follows that 
$\lim_{T\rightarrow\infty}\gamma^T_i=\lambda_i$. A large deviation approach
suggests that the probability density function (PDF) of the largest 
stretching rate $\gamma_1$ measured over a long time 
$T\gg 1/\lambda_1$ takes the asymptotic form  
$P_{T}(\gamma_1)\sim N(t)\exp[-H(\gamma_1)T]$ where the Cram\'er function 
$H(\gamma_1)$ is convex and obeys the conditions 
$H(\lambda_1)=0$, $H^\prime(\lambda_1)=0$. 
We computed the Cram\'er function for the Lagrangian FTLE for the Newtonian case
and the two viscoelastic cases. 
In the inset of Fig.~\ref{fig4} we plotted the average of the stretching rates
(i.e., the first Lagrangian Lyapunov exponent of the flow $\lambda_1^0$) and
the rescaled variance $\mu=T\langle\gamma_1^2\rangle$, for the three values of
$\rm Wi$ that we considered.  The decrease of the Lyapunov exponent (rescaled
with the Newtonian value $\lambda_1^N$ for comparison) gives a measure of the
decrease in the chaoticity of the flow, due to the action of Polymers. On the
other hand, we also observe a decrease in the relative variance
$\mu/\Lambda_1^0$, which implies that polymer feedback induces also a reduction
of the fluctuations of of stretching rates. Inspection of the main panel of
Fig.~\ref{fig4}, however, shows that fluctuations are not reduced uniformly.
Indeed, the shape of $P(\gamma_1)$ changes when polymers are added.
As is evident in Fig.\ref{fig4}, elasticity has the effect of raising the 
right branch of the Cram\'er function, while the left one is comparatively 
less affected. 
Given the definition of $H(\gamma_1)$, this amounts to a relative
suppression of positive fluctuations in the stretching rate: as one could
expect, polymers have a larger (negative) feedback on events of larger
stretching.

\begin{figure}
\includegraphics[width=8.0cm]{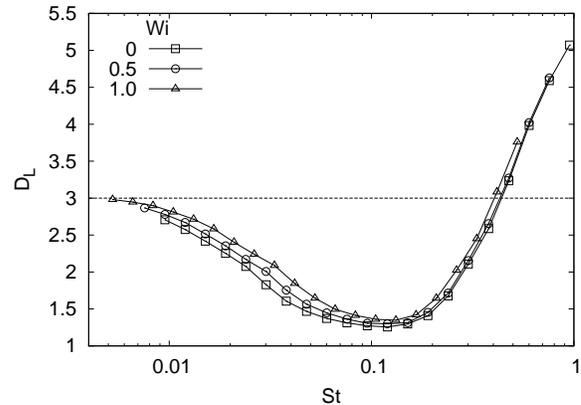}
\includegraphics[width=8.0cm]{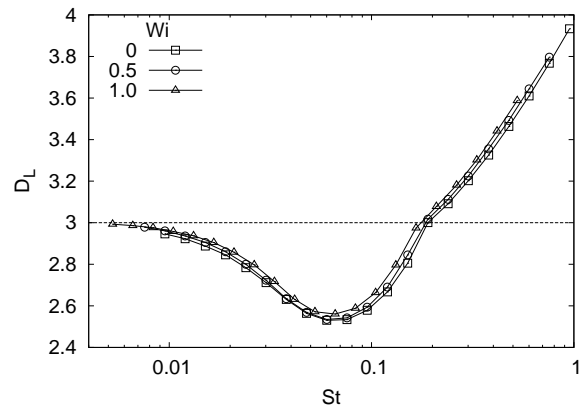}
\caption{
Lyapunov dimension for light (upper panel) and heavy (lower panel) particles
plotted as a function of ${\rm St=\tau_S \lambda^0_1}$. Different lines correspond
to the different Weissenberg numbers with symbols as in Fig.~\ref{fig2}.
}
\label{fig5}
\end{figure}

The effect of polymers on Lyapunov exponents and the Lagrangian nature of the latter suggests to  introduce the dimensionless Stokes
number defined as $\rm St=\tau_S \lambda^0_1$ which depends on $\rm Wi$ by the
dependence of $\lambda^0_1$ shown in Fig.~\ref{fig4}. Figure~\ref{fig5} 
shows the Lyapunov dimension $D_L$ for both heavy and light particles 
as a function of $\rm St$.
It is evident that, with respect to Fig.~\ref{fig2},
the collapse of the curves at different $\rm Wi$ is improved. 
In particular, the minimum 
of the fractal dimension (which corresponds to maximum clustering) occurs 
almost for the same $\rm St$ number. 
Still, some differences are observable, in particular for small $\rm St$ in the case
of light particles. 
This can be understood by the following argument. 
Bubbles, at variance with heavy particles, 
have tendency to concentrate on filaments of high vorticity. 
Indeed, while the minimal dimension for heavy particles is about $2.5$
(at $\rm St \simeq 0.1$), for light particles at maximal clustering 
it becomes as small as $1.26$.
Vortex filaments correspond to quasi-one-dimensional
regions of intense stretching, in the direction
longitudinal to the vortex, which give major contributions to the right
tail of the Cram\'er function. 
As shown in Fig.~\ref{fig4}, the effects of polymers on the distribution
of Lyapunov exponent is more evident in this region of strong fluctuations,
where the distribution does not rescale with $\lambda^0_1$. It is therefore
not surprising that also the effects on clustering of light particles
cannot be completely absorbed in the rescaling of $\tau_S$ with the mean
stretching rate $\lambda^0_1$.

\begin{figure}
\includegraphics[width=8.0cm]{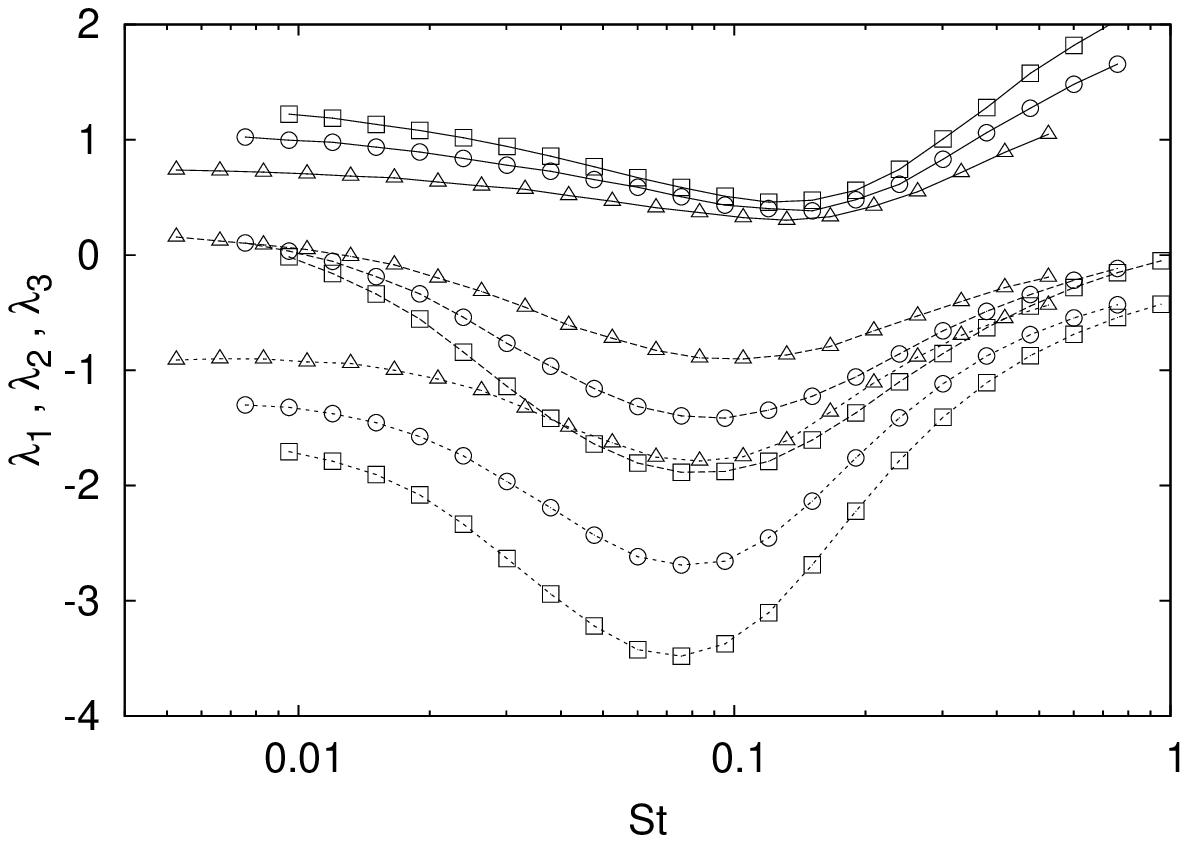}
\includegraphics[width=8.0cm]{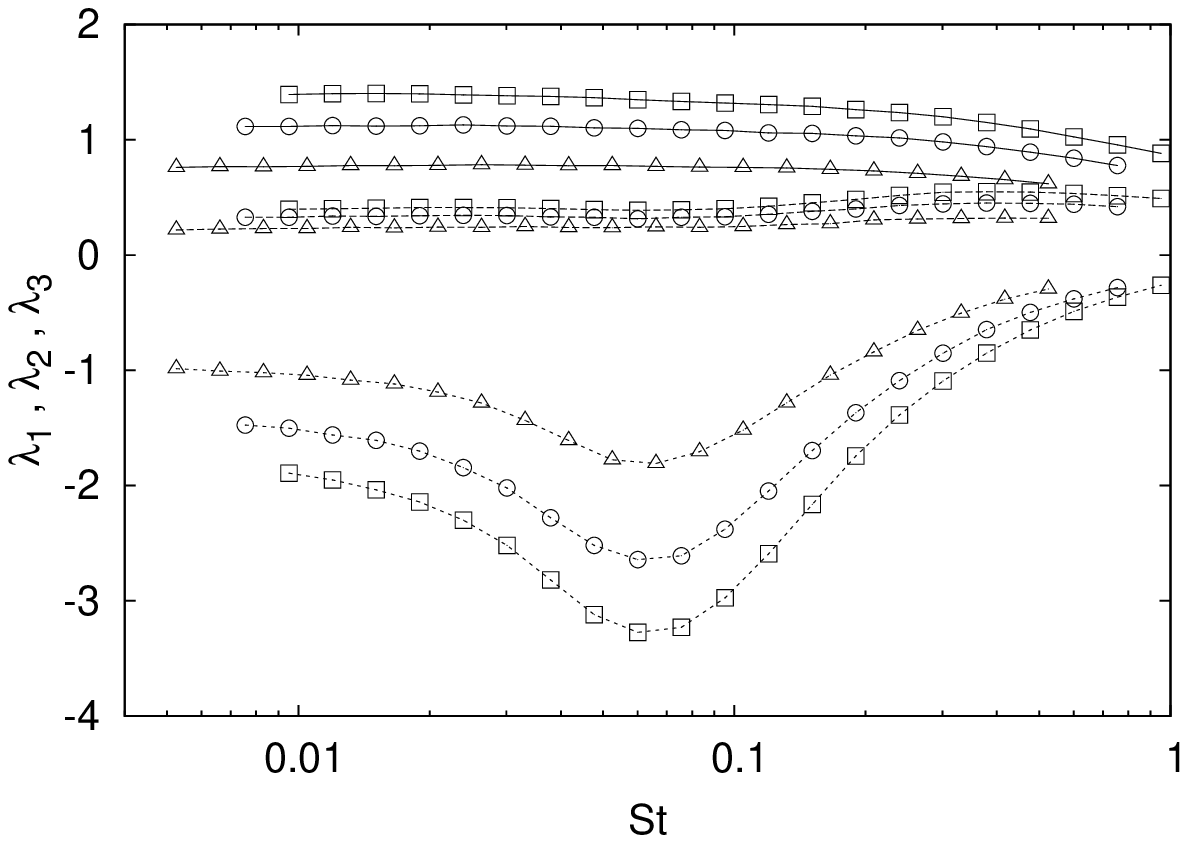}
\caption{The first three Lyapunov exponents for light ($\beta=3$, upper panel)
and heavy ($\beta=0$, lower panel) particles, at different $\rm Wi$.
Continuous, dashed and dotted lines represent the first, second and
third Lyapunov exponents, while symbols correspond to different $\rm Wi$ 
as in Fig.~\ref{fig2}.
}
\label{fig6}
\end{figure}

As the fractal dimension is given by a combination of the Lyapunov exponents,
in order to better understand the differences on light and heavy particles,
in Fig.\ref{fig6} we show the first three Lyapunov exponents as a function
of $\rm St$. The first observation is that bubbles, at variance with heavy 
particles, exhibit negative values of $\lambda_2$, consistently with the 
lower value of $D_L$ and the tendency of light particles to concentrate 
towards vortex filaments.

The first Lyapunov exponent decreases with $\rm Wi$ for any value of $\rm St$,
thus indicating that the phenomenon of chaos reduction, already discussed 
for the case of Lagrangian tracers, is generic also for inertial particles.  
On the contrary, the second Lyapunov exponent shows a different behavior
for light and heavy particles: it increases for the former but slightly 
decreases for the latter. Figure~\ref{fig6} shows that the effect of 
polymers is not a simple rescaling of the Lyapunov spectrum, which 
would trivially keep the dimension $D_L$ unchanged. From this point
of view, the almost perfect rescaling of the Lyapunov dimensions
shown in Fig.~\ref{fig5} is quite surprising and arises as the result
of compensations of different effects.

In conclusion, we investigated the clustering properties of inertial 
(heavy and light) particles in a turbulent viscoelastic fluid. 
The main effect of polymers on turbulent
flows is to counteract small-scale fluctuations and to reduce its chaoticity.
Quantitatively, this results in a decrease in the first Lyapunov exponent of 
the flow, which, in turn, affects clustering of inertial particles. 
The latter can be quantified by means of the fractal (Lyapunov) dimension of
particle distributions. Although the effects of polymers on the particle
Lyapunov exponents are complex and qualitatively different for light 
and heavy particles, the overall effect on fractal dimension is relatively
simple and can be rephrased in the rescaling of the characteristic
time of the flow. 
Indeed, when particle inertia is parametrized by the Stokes number $\rm St$ 
defined with the Lyapunov time of the flow, one can approximately rescale 
the curves $D_L(\rm St)$ at all $\rm Wi$. 
In contrast, as polymers do not affect large scale properties of the 
flow, a parametrization of particle inertia based on integral time scales
would not show a collapse of the curves $D_L(\rm St)$ at different $\rm Wi$. 
As a consequence, any prediction of
particle clustering in turbulent polymeric solutions requires an accurate
estimate of small scale stretching rates.

We acknowledge support from the the EU COST Action MP0806.


\end{document}